\documentclass[11pt]{article}

\usepackage{xcolor}
\usepackage{latexsym}
\usepackage{amsbsy}
\usepackage{amssymb}
\usepackage{amsmath}
\usepackage{shadow}
\usepackage{fancybox}
\usepackage{pifont}
\usepackage{graphics}
\usepackage{cite}
\usepackage{rotating}
\usepackage{color, colortbl}

\def\0{\mbox{\tiny $0$}}
\def\1{\mbox{\tiny $1$}}
\def\2{\mbox{\tiny $2$}}
\def\3{\mbox{\tiny $3$}}
\def\4{\mbox{\tiny $4$}}
\def\5{\mbox{\tiny $5$}}
\def\6{\mbox{\tiny $6$}}
\def\7{\mbox{\tiny $7$}}
\def\8{\mbox{\tiny $8$}}
\def\9{\mbox{\tiny $9$}}

\def\rvec{\hspace{-1pt}(\mathbf{r})}
\def\A{\mathcal{A}}
\def\hyd{_{_{\textrm{hyd}}}}
\def\osc{_{_{\textrm{osc}}}}
\def\well{_{_{\textrm{well}}}}
\def\rel{_{_{\textrm{rel}}}}

\def\L{_{_{\textrm{L}}}}

\newcommand{\fpsi}[2][r]{\psi_{{#2}}\hspace{-1pt}(\mathbf{#1})}
\newcommand{\fphi}[2][r]{\varphi_{{#2}}\hspace{-1pt}(\mathbf{#1})}

\newcommand{\dagphi}[2][r]{\varphi_{{#2}}^{\dagger}(\mathbf{#1})}
\newcommand{\qE}{\mathcal{E}}

\newcommand{\conj}[1]{{#1}^{^{\star}}}

\newcommand{\rint}{\int\textrm{d}\mathbf{r}\,\,}
\newcommand{\abs}[1]{\lvert #1 \rvert}

\begin{document}

\date{\normalsize \textbf{\color{red!60!black}{European Physics Journal Plus}} \textbf{134}, 113-125 (2019). }
\title{\hspace*{-.5cm}\shadowbox{\fcolorbox{black}{gray!50} { {\color{blue!60!black}{ \Large \textbf{
\begin{tabular}{c}
QUATERNIONIC PERTURBATION THEORY
\end{tabular}}}}}}
}

\author{
{\large \textbf{\color{red!60!black}{ Stefano De Leo}}}\\
{\normalsize Applied Mathematics Department,}\\
{\normalsize State University of Campinas,}\\
{\normalsize  Campinas (SP), Brazil}\\
{\normalsize  \textit{\color{blue!60!black}{deleo@ime.unicamp.br}}}\\
 \and
{\large \textbf{\color{red!60!black}{ Gisele Ducati}}}\\
{\normalsize CMCC,}\\
{\normalsize Federal University of ABC,}\\
{\normalsize  Santo Andr\'e (SP), Brazil}\\
{\normalsize \textit{\color{blue!60!black}{ducati@ufabc.edu.br}}}\\
 \and
{\large \textbf{\color{red!60!black}{ Caio Almeida Alves de Souza}}}\\
{\normalsize CCNH,}\\
{\normalsize Federal University of ABC,}\\
{\normalsize  Santo Andr\'e (SP), Brazil}\\
{\normalsize \textit{\color{blue!60!black}{caio.almeida.ads@gmail.com}}}
}

\maketitle

\abstract{In this paper we present a perturbation theory for constant quaternionic potentials. The effects of quaternionic perturbations  are explicitly treated for bound states of hydrogen atom, infinite potential well and harmonic oscillator. Comparison with relativistic corrections is also briefly discussed.}





\section*{\large \color{blue!60!black}{I. INTRODUCTION}}
\qquad In the last years, many papers investigated qualitative and quantitative differences between the complex and the quaternionic  formulation of quantum mechanics \cite{del1,del2,del3,del4,del5,del6,del7,del8,art1,art2}. Since square potentials are often used to illustrate implications of quantum mechanics it is natural, when passing from the complex to the quaternionic formalism, to re-examine  such kind of potentials by extending their nature from complex to quaternions. A detailed discussion of quaternionic formulations of quantum theories is found in the valuable book of Adler \cite{adler}.

 In comparing complex with quaternionic quantum mechanics,  the effects of quaternionic potentials have often been evaluated by considering from  the full quaternionic soilution its small perturbation limit. The work presented in this paper was  motivated by the following consideration: If quaternionic potentials are treated as perturbation effects of complex quantum mechanics, could we calculate the energy eigenvalues modification by using a quaternionic perturbation theory? In this way, we do not need to find the explicit quaternionic solution but only the quaternionic perturbation on the standard (complex) analysis. Exactly in the same way in which relativistic corrections can be obtained from the energy eigenvalues solutions of the Schr\"odinger equation \cite{cohen} without solving the Dirac equation \cite{zuber}.

Observing that in physics we measure energy differences and  not absolute energies, we calculate, starting by some typical solution  of  complex quantum mechanics,  the effect that constant quaternionic perturbations play on the energy gap between bound states. It is well known \cite{cohen} that constant complex potentials do not affect the calculation of the energy gap. If quaternionic potential  affects such a calculation, we immediately have a \textit{qualitative} difference between the two formulation. If this happens, we can also calculate, by using a revised  perturbation theory, the quaternionic perturbation effects and this gives us the possibility to also see \textit{quantitative} difference between the complex and quaternionic formulation and, consequently,  to evaluate the order of magnitude of these quaternionic effects in view of possible experimental confirmations.

In the next section, we fix our notation and introduce hydrogen atom bound states of complex quantum mechanics. In section III, we present the quaternionic perturbation theory and then apply it to some specific cases  in section IV.
In such a section, we also   compare the quaternionic perturbation with relativistic corrections. Our conclusion are drawn in the final section.

\section*{\large \color{blue!60!black}{II. BOUND STATES IN COMPLEX QUANTUM MECHANICS}}
Let us consider the time-independent (complex) Schr\"odinger equation \cite{cohen}
\begin{equation}
H \, \fphi{} = E \, \fphi{}\,\,,
\end{equation}
with $E\in\mathbb{R}$, $\fphi{}: \mathbb{R}^{3} \to \mathbb{C}$  and  where
\[ H = - \dfrac{\hbar^{^2}}{2\,m} \nabla^{^2} + V\rvec\,\,,\]
with $\nabla \equiv \left(\partial_{x}, \partial_{y}, \partial_{z} \right)$ and
$V\rvec:\mathbb{R}^{3} \to \mathbb{R}$.

In order to fix our notation an introduce the concept of bound state, we introduce a well known
physical system containing an interaction potential, i.e. the hydrogen atom.  One proton ($m_{p}$),
one electron ($m_e$), and the electrostatic (Coulomb) potential that holds them together,
\[V\hyd\rvec = -\,\frac{e^{^{2}}}{4\,\pi\,\epsilon_{_{0}}\,r}\,\,,\]
where $r=\sqrt{x^{^{2}}+ y^{^{2}} +z^{^{2}} }$ is the distance between the proton ($+e$) and electron ($-\,e$) charges and $\epsilon_{\0}$ the vacuum permittivity.

Ignoring all spin-coupling interaction and using the reduced mass
\[\mu=m_em_p/(m_e+m_p)\,\,,\]
the hydrogen Schr\"odinger equation,
\begin{equation}
\left[\, - \dfrac{\hbar^{^2}}{2\,\mu} \nabla^{^2} + V\hyd\rvec\right]\varphi\hyd(\mathbf{r})=E\hyd\varphi\hyd(\mathbf{r})\,\,,
\end{equation}
can be solved, after expanding the Laplacian in spherical coordinates, by separation of variables
leading to  eigenfunctions containing  spherical harmonic and generalized Laguerre special functions\cite{cohen}. The energy eigenvalues are given by
\begin{equation}
E\hyd(n)= -\, \frac{\mu\,e^{^{4}}}{2\,(\,4\,\pi\,\epsilon_{_{0}})^{^{2}}\hbar^{^{2}}}\,\,\frac{1}{n^{^{2}}}=-\,
\frac{R_y}{n^{^{2}}}\,\,,
\end{equation}
where $R_y\approx13.6\,\mathrm{eV}$ is the Rydberg unit of energy.

The spectral frequency condition, also known as Rydberg formula, is the given by
\begin{equation}
E\hyd(m)-E\hyd(n)=-\,\left(\,\frac{1}{m^{^{2}}}\,-\,\frac{1}{n^{^{2}}}\,\right)\,R_y\,\,,
\end{equation}
which in the special case of $m=n+1$ reduces to
\begin{equation}
\Delta E\hyd(n)=E\hyd(n+1)-E\hyd(n)=\frac{2\,(n+1)}{n^{^{2}}(n^{^{2}}+1)}\,R_y\,\,.
\end{equation}
How a constant quaternionic perturbation changes the spectral frequencies is one of the objective of this article.

\section*{\large \color{blue!60!black}{III. QUATERNIONIC PERTURBATION THEORY }}

In quaternionic quantum mechanics, we deal with antihermitian operators and with their complex right eigenvalues \cite{right1,right2}. The effects of  quaternionic potentials  have been often interpreted as small perturbations of the results obtained in complex quantum mechanics \cite{del1,del2,adler}. Even though this treatment as a perturbation  has been  commented, it appears that a quaternionic perturbation theory has not been yet discussed.
 Perturbation mathematical methods are used in physics to find   approximate solutions to a given problem  by starting from the  exact solution of a, related,  simpler problem.   Several physical systems are indeed described by equations that are too complicated to be solved analytically. Adding quaternionic terms to complex potentials, we do not expect things to get easier. The possibility to use a quaternionic perturbation theory could allow to find approximate solutions of a major real of problems and get an immediate comparison with their complex counterpart. Formulating a quaternionic prturbation theory is the main objective of this paper.

In the presence of a constant pure quaternionic  potential $j\,\alpha\,W$,
\[\alpha \in \mathbb{R}\,\,,\,\,\,\,\,W\in \mathbb{C}\,\,,\,\,\,\,\, \mathrm{and}\,\,\,\,\,i^{\2} = j^{\2} = k^{\2} = ijk = -1\,\,,\]
  we should to solve the Schr\"odinger equation \cite{adler}
\begin{equation} \label{qse}
\A \, \Phi\rvec = \qE \, \Phi\rvec \, i\,,
\end{equation}
where the quaternionic anti-hermitian operators, whose right (complex) eigenvalues  \cite{right1,right2} lead to the energy solutions, is given by
\[
\A = i\,H + j\,\alpha\,W\,\,.
\]
Observe that  for the probabilistic essence of quantum mechanics, the associativity (and not the commutativity) is the  crucial attribute to guarantee an adequate probability interpretation \cite{adler}. Consequently, amplitudes of probabilities can only be properly defined in \textit{associative division algebras} and quaternions are the first and the last possible
extension of complex maintaining the property of an associative division algebra.

Perturbation theory leads to an expression for the desired solution in terms of a formal power series in some small parameter, in our case $\alpha$, that quantifies the deviation from the exactly solvable problem. The leading term in this power series is the solution of the complex exactly solvable problem, while further terms describe the quaternionic deviation in such a solution.
 To shorten our notation, we make use of the Einstein summation convention over the
repeated indices. In this formalism, the eigenfunctions can be rewritten in terms of the complex eigenfunction $\fphi{}$ and its higher complex and quaternionic terms corrections,
\begin{equation} \label{wfp}
\Phi\rvec = \fphi{} + \alpha^{s} \fphi{s} + j \hspace{1pt} \alpha^{s} \fpsi{s}\, \,,
\end{equation}
where $\fphi{s}, \fpsi{s}: \mathbb{R}^{3} \to \mathbb{C}$ and $s=1,2,...$. The (real) energy eigenvalue $\qE$ is then expressed in terms of the  $s$-th order corrections by
\begin{equation} \label{ep}
\qE = E + \alpha^{s} E_{s}\,\,.
\end{equation}
The quaternionic Schr\"odinger equation can be rewritten as two coupled complex equations. The first  equation comes from the complex part of Eq.(\ref{qse}),
\begin{equation}
\label{cpart}
\alpha^{s}\,H\, \fphi{s}\,+\,i\,\alpha^{s\mbox{\tiny{+1}}}\,\conj{W}\fpsi{s}\, =\, \alpha^{s}\,E\,\fphi{s} \,+\,\alpha^{s}\,E_s\,\fphi{} \,+\,\alpha^{s\mbox{\tiny{+}}u}\,E_s\,\fphi{u}\,\,,
\end{equation}
and the second coupled equation from the pure quaternionic part of Eq.(\ref{qse}),
\begin{equation}
\label{qpart}
-\,i\,\alpha\,W\,\fphi{}\,-\,i\,\alpha^{s\mbox{\tiny{+1}}}\,W\,\fphi{s}\,-\,\alpha^{s}\,H\,\fpsi{s} =\, \alpha^{s}\,E\,\fpsi{s} \,+\,\alpha^{s\mbox{\tiny{+}}u}\,E_s\,\fpsi{u}\,\,.
\end{equation}

\subsection*{\normalsize \color{blue!60!black}{III.A FIRST ORDER CORRECTION}}

Let us begin our analysis by calculating the first order correction to the energy eigenvalues of complex quantum mechanics. Equation (\ref{cpart}) reduces to
\begin{equation}
H\, \fphi{\1}\, = \, E\,\fphi{\1}\,+\,E_{\1}\,\fphi{}\,\,,
\end{equation}
which multiplying by $\dagphi{}$ from the left and integrating them over $\boldsymbol{r}$
becomes
\begin{equation}
\label{cpart1}
 \rint \dagphi{}\,H\,\fphi{1}\, =\, E \,\rint \dagphi{}\, \fphi{1}  \,+\,N\, E_{\1} \,\,,
  \end{equation}
where $N$ is the normalization constant of the complex solution $\fphi{}$. Observing that
\[\rint \dagphi{}\,H\,\fphi{1} \,=\,  \rint \big[ H \fphi{} \big]^{\dagger} \fphi{1}\, =\,E\, \rint \dagphi{} \,\fphi{1}\,\,,  \]
from equation (\ref{cpart1}) we immediately get
\begin{equation} \label{E1A}
E_{\1} = 0 \,.
\end{equation}
 Equation (\ref{qpart}) reduces to
\begin{equation}
-\,i\,W\,\fphi{}\,-\,H\,\fpsi{\1}\,=\,E\,\fpsi{\1}\,\,.
\end{equation}
Repeating what done for equation (\ref{cpart}), we obtain
\begin{equation} \label{E1B}
i\,\conj{W}\rint \dagphi{} \fpsi{1} = N\, \frac{\,\,\abs{W}^{^{2}}}{2\,E} \,\, .
\end{equation}
This equation will be used in the next section to calculate the second order correction of the energy eigenvalue

\subsection*{\normalsize \color{blue!60!black}{III.B SECOND ORDER CORRECTION}}

Let us now consider the second order term of the correction.  By using $E_{\1}=0$, from equation (\ref{cpart}) we find
\begin{equation}
H\,\fphi{\2} \,+\,i\, \conj{W} \fpsi{\1} \,=\, E \,\fphi{\2} + E_{\2} \,\fphi{} \,\,.
\end{equation}
As done in the previous subsection, this equation can be converted in an integral equation
\begin{equation}
\label{cpart2}
 \rint \dagphi{}\,H\,\fphi{\2} \,+\,i\, \conj{W}\,\rint \dagphi{}\,\fpsi{\1} \,=\, E \rint \dagphi{} \fphi{\2}  \,+\,N\, E_{\2} \,\,,
  \end{equation}
from which, by using the relation (\ref{E1B}), we obtain the second order correction to the energy eigenvalue $E$,
 \begin{equation} \label{E2A}
E_{\2} = \frac{\,\,\abs{W}^{^2}}{2\,E}\,\,.
\end{equation}
The second equation (\ref{qpart}) will be once again employed to calculate an integral  used later in the calculation of the higher order term in the energy eigenvalue expansion,
\begin{equation}
\label{qpart2}
 -\,i\,W\,\rint \dagphi{}\,\fphi{\1} \,-\,\rint \dagphi{}\,H\, \fpsi{\2}
 =\, E \rint \dagphi{} \fpsi{\2}\,\,.
  \end{equation}
After simple algebraic manipulation, we find
\begin{equation} \label{E2B}
i\,\conj{W}\rint \dagphi{} \fpsi{\2}  = E_{\2}\, \rint \dagphi{} \fphi{\1}\,\, .
\end{equation}

\subsection*{\normalsize \color{blue!60!black}{III.C THIRD ORDER CORRECTION}}

For the third-order energy correction, by using the first (\ref{E1A}) and second (\ref{E2A}) expansion terms, from equation (\ref{cpart}) we obtain
\begin{equation}
H\,\fphi{\3} \,+\,i\, \conj{W} \fpsi{\2} \,=\, E \,\fphi{\3} \,+\, E_{\2}\,\fphi{\1}
\,+\, E_{\3} \,\fphi{} \,\,.
\end{equation}
The correspnding integral equation
\begin{equation}
\label{cpart3}
 i\, \conj{W}\,\rint \dagphi{}\,\fpsi{\2} \,=\,E_{\2}\,\rint \dagphi{}\fphi{\1}     \,+\,N\, E_{\3}\,\,,
  \end{equation}
by using the realtion (\ref{E2B}), immediatly leads to
\begin{equation}
E_{\3}=0\,\,.
\end{equation}
The third order expansion of the intergal equation  coming from (\ref{qpart}) gives
\begin{equation}
\label{qpart3}
 -\,i\,W\,\rint \dagphi{}\,\fphi{\2} \,
 =\, 2\,E \rint \dagphi{} \fpsi{\3}\,+\,E_{\2} \rint \dagphi{} \fpsi{\1}\,\,,
  \end{equation}
from which by using (\ref{E1B}) and   (\ref{E2A}) we get
\begin{equation} \label{E3B}
i\,\conj{W}\rint \dagphi{} \fpsi{\3} \,=\, E_{\2} \rint \dagphi{} \fphi{\2}\, -\, N\,\frac{\,\,E_{\2}^{^{2}}}{2\,E}\,\,.
\end{equation}

\subsection*{\normalsize \color{blue!60!black}{III.D FOURTH AND SIXTH ORDER CORRECTIONS}}
Before looking for a close expression for the energy eigenvalues expansion let us perform the last explicit calculations, useful in understanding the higher terms correction and important in finding recurrence  relations. The integral equation coming from the fourth order term in equation (\ref{cpart}) is given by
\begin{equation}
\label{cpart4}
 i\, \conj{W}\,\rint \dagphi{}\,\fpsi{\3} \,=\,E_{\2} \,\rint \dagphi{}\fphi{\2}     \,+\,N\, E_{\4} \,\,.
  \end{equation}
By using  the realtion (\ref{E3B}), we obtain
\begin{equation} \label{E4A}
E_{\4} = - \,\frac{\,\,E_{\2}^{^2}}{2\,E}\,\,.
\end{equation}
The integral equation for the sixth order is
\begin{equation}
\label{cpart6}
 i\, \conj{W}\,\rint \dagphi{}\,\fpsi{\5} \,=\,E_{\2} \,\rint \dagphi{}\fphi{\4}     \,+\,
 E_{\4} \,\rint \dagphi{}\fphi{\2}     \,+\,N\, E_{\6} \,\,.
  \end{equation}
To calculate the energy eigenvalue, we have to use the integral equation coming from the fifth order term in equation (\ref{qpart}), i.e.
\[
 -\,i\,W\,\rint \dagphi{}\,\fphi{\4} \,
 =\, 2\,E \rint \dagphi{} \fpsi{\5}\,+\,E_{\2} \rint \dagphi{} \fpsi{\3}\,+\,
 E_{\4} \rint \dagphi{} \fpsi{\1}
 \,\,,
\]
which after some algebraic manipulation can be rewritten as follows
\begin{eqnarray}
i\,\conj{W} \rint \dagphi{} \fpsi{\5} & = &  E_{\2} \rint \dagphi{} \fphi{\4} \,-\,i\conj{W}
\frac{E_{\2}}{2\,E}\,\rint \dagphi{} \fpsi{\3}\,-\,i\,\conj{W} \frac{E_{\4}}{2\,E}\,\rint \dagphi{} \fpsi{\1}\nonumber\\
& = & E_{\2} \rint \dagphi{} \fphi{\4} \,+\,E_{\4} \rint \dagphi{} \fphi{\2}\,-\,N\,\frac{E_{\2}E_{\4}+E_{\4}E_{\2}}{2\,E}\,\,.
\end{eqnarray}
Substituting the previous equation in (\ref{cpart6}), we obtain
\begin{equation}
E_{\6}=-\,\frac{\,\,E_{\2}\,E_{\4}}{E}\,\,.
\end{equation}

\subsection*{\normalsize \color{blue!60!black}{III.E CLOSED FORMULA}}
Let us now generalize the results found in the previous subsection, considering for example the first ten terms in the energy eigenvalues expansion
\begin{eqnarray}
2\,E\,\left\{\,E_{\2}\,,\,E_{\4}\,,\,E_{\6}\,,\,E_{\8}\,,\,E_{\1\0}\,\right\} & = & \left\{\,
\abs{W}^{^{2}}\,,\,-\,E_{\2}^{^{2}}\,,\,-\,2\,E_{\2}\,E_{\4}\,,\,-\,2\,E_{\2}\,E_{\6}-E_{\4}^{^{2}}\,,\,
-\,2\,(\,E_{\2}\,E_{\8}+E_{\4}\,E_{\6}\,)\,\right\}\nonumber \\
 & = & \left\{\,
\abs{W}^{^{2}}\,,\,-\,E_{\2}^{^{2}}\,,\,\frac{E_{\2}^{^{3}}}{E}\,,\,
-\,\frac{5\,E_{\2}^{^{4}}}{4\,E^{^{2}}}\,,\,\frac{7\,E_{\2}^{^{5}}}{4\,E^{^{3}}}\,\right\}\nonumber\\
 & & \left\{\,
\abs{W}^{^{2}}\,,\,-\,\frac{\abs{W}^{^{4}}}{4\,E^{^{2}}}\,,\,\frac{\abs{W}^{^{6}}}{8\,E^{^{4}}}\,,\,
-\,\frac{5\,\abs{W}^{^{8}}}{64\,E^{^{6}}}\,,\,\frac{7\,\abs{W}^{^{8}}}{128\,E^{^{6}}}\,\right\}\,\,.
\end{eqnarray}
To find the closed formula for the energy eigenvalues it is convenient reorganize the  previous coefficients in the following form
\[
\left\{\,\frac{2\,E\,E_{\2}}{\abs{W}^{^{2}}}\,,\,2\,\frac{(2\,E)^{^{3}}\,E_{\4}}{\abs{W}^{^{4}}}\,,\,
3\,\frac{(2\,E)^{^{5}}\,E_{\6}}{\abs{W}^{^{6}}}\,,\,4\,\frac{(2\,E)^{^{7}}\,E_{\8}}{\abs{W}^{^{8}}}
\,,\,5\,\frac{(2\,E)^{^{9}}\,E_{\1\0}}{\abs{W}^{^{10}}}\,,\,...\,,\,
s\,\frac{(2\,E)^{^{2\,{\mbox{\tiny $s$}\,-\,1}}}\,E_{\2 s}}{\abs{W}^{^{2{\mbox{\tiny $s$}}}}}
\,,\,...\,\right\}
\]
 obtaining the following sequence
\[  \left\{\,1\,,\,-\,2\,,\,6\,,\,-\,20\,,\,70\,,\,...\,,\, (-\,1)^{^{\mbox{\tiny $s$\,+\,1}}}
\left(\begin{array}{r} 2\,s-2\\s-1\end{array}
\right)  \,,\,...\,\right\}\,\,.\]
Finally, the energy eigenvalues series is
\begin{equation} \label{Eseries}
\frac{\qE}{E} = 1 + \sum_{s=1}^{\infty} (-\,1)^{^{\mbox{\tiny $s$\,+\,1}}} \, \dfrac{2}{s}
\, \left(\begin{array}{r} 2\,s-2\\s-1\end{array}
\right)\,
 \left( \dfrac{\alpha\, \abs{W}}{2\,E} \right)^{^{2\,\mbox{\tiny $s$}}} \,\, .
\end{equation}
An alternating series converges if two conditions are met \cite{Eseries}:
the convergence to zero in the limit $s\to \infty$ of the $s$-th term
and if each term is either smaller than or the same as its predecessor (ignoring the minus signs).
So the series (\ref{Eseries}) converges for
\[
\lim_{s\,\to\, \infty} \displaystyle{\frac{\dfrac{2}{s+1}
\, \left(\begin{array}{r} 2\,s\\s\end{array}
\right)\,\left( \dfrac{\alpha\, \abs{W}}{2\,E} \right)^{^{2\,\mbox{\tiny $s$}\,+\,2}}
}{\dfrac{2}{s}
\, \left(\begin{array}{r} 2\,s-2\\s-1\end{array}
\right)\,\left( \dfrac{\alpha\, \abs{W}}{2\,E} \right)^{^{2\,\mbox{\tiny $s$}}}}}\,\,\leq\,\, 1\,\,\,\,\,\,\Rightarrow\,\,\,\,\,\,\, \abs{\alpha\,W}\,\leq\,\abs{\,E\,}\,\,.
\]

\section*{\large \color{blue!60!black}{IV. APPLICATIONS}}

In this section, we apply the results obtained for quaternionic perturbations to well-known  quantum systems like hydrogen atom, potential well and harmonic oscillator. But we shouldn't only introduce the energy eigenvalues in the series \eqref{Eseries} and see what is the outcome. If we are going to study the quaternionic deviations of standard (complex) systems, we should describe them in terms of a measurable quantity. In physics we measure energy differences not absolute energies,  so this is precisely what we are going to calculate throughout this section: quaternionic perturbations on the energy difference between two  quantum levels.

\subsection*{\normalsize \color{blue!60!black}{IV.A HYDROGEN ATOM}}

Since the series (\ref{Eseries}) is completely general in respect to what we defined as the starting  quantum system, then, as a first example, we apply our results to one of the most studied quantum system, i.e. the hydrogen atom. In the second section, we briefly discussed the Ryderberg formula for the spectral frequencies. What we aim in this subsection is to evaluate the quaternionic perturbations of the adimensional Ryderberg formula
\begin{equation}
\lambda\hyd(n)=\frac{\Delta E\hyd(n)}{R_y}=\frac{E\hyd(n+1)-E\hyd(n)}{R_y}=
\frac{2\,n+1}{n^{^{2}}(n+1)^{^{2}}}\,\,.
\end{equation}
In order to simplify the presentation, it is convenient to express the quaternionic perturbation in terms of the Ryderberg energy, $\abs{W}=2\,R_y$.  By using the energy series (\ref{Eseries}), we obtain
\begin{equation}
\frac{\qE\hyd(n,\alpha)}{R_y} = -\,\frac{1}{n^{^{2}}}\, -\,\frac{1}{n^{^{2}}}\, \sum_{s=1}^{\infty} (-\,1)^{^{\mbox{\tiny $s$\,+\,1}}} \, \dfrac{2}{s}
\, \left(\begin{array}{r} 2\,s-2\\s-1\end{array}
\right)\,
 \left(n^{\2}\,\alpha \right)^{^{2\,\mbox{\tiny $s$}}}\,\,,
\end{equation}
which, by recalling the convergence condition $\alpha\,W\,/\,E(n)\,|\leq1$, represents a convergent
series for
\begin{equation}
\alpha \,\,\leq \,\,\alpha\hyd\,\,=\,\,\frac{1}{2\,n^{^{2}}}\,\,.
\end{equation}
The quaternionic Ryderber formula is then given by
\begin{equation} \label{gap_Q}
\Lambda\hyd(n,\alpha)=\frac{\Delta \qE\hyd(n,\alpha)}{R_y}=\frac{\qE\hyd(n+1,\alpha)-\qE\hyd(n,\alpha)}{R_y}\,\,.
\end{equation}
We can estimate the correction to the (complex) spectral frequencies by introducing the following $\sigma$ ratio,
\begin{eqnarray}
\sigma\hyd(n,\alpha)&=&
\Lambda\hyd(n,\alpha)\,/\,\lambda\hyd(n)\nonumber\\
 & = & 1\,+\, \frac{n^{^{2}}(n+1)^{^{2}}}{2\,n+1}\,\sum_{s=1}^{\infty} (-\,1)^{^{\mbox{\tiny $s$}}} \, \dfrac{2}{s}
\, \left(\begin{array}{r} 2\,s-2\\s-1\end{array}
\right)\,\left[\,(n+1)^{^{4\,\mbox{\tiny $s$}\,-\,2}} - n^{^{4\,\mbox{\tiny $s$}\,-\,2}}  \,\right]
 \alpha^{^{2\,\mbox{\tiny $s$}}}\,\,.
\end{eqnarray}
In Fig.\,1, we plot $\sigma\hyd(n,\alpha)$ as a function of the $s$-th order correction term for different (convergent) values of $\alpha$. For $n=1$, the convergence is guarantee for $\alpha\leq 1/8$, see Fig.\,1(a). For $n=2$ the upper (convergence) limit is given by $1/18$, see Fig.\,1(b).

\subsection*{\normalsize \color{blue!60!black}{IV.B POTENTIAL WELL}}

The energy eigenvalues for a particle trapped in a one-dimensional infinite potential well are given by \cite{cohen},
\begin{equation}
E\well(n) = n^{\2}\,E\L\,\,,
\end{equation}
where $E\L=\pi^{\2} \hbar^{^2}\,/\,2\, m_e\,L^{^{2}}$ with $L$ indicating the width of the infinite well. As done in the previous subsection, we aim to calculate the effects of quaternionic perturbation on
\begin{equation}
\label{wellC}
\lambda\well(n)=\frac{\Delta E\well(n)}{E\L}=\frac{E\well(n+1)-E\well(n)}{E\L}=
2\,n+1\,\,.
\end{equation}
By using $\abs{W}=2\,E\L$, the energy series (\ref{Eseries}) reduces to
\begin{equation}
\frac{\qE\well(n,\alpha)}{E\L} = n^{\2}\, +\,n^{\2}\, \sum_{s=1}^{\infty} (-\,1)^{^{\mbox{\tiny $s$\,+\,1}}} \, \dfrac{2}{s}
\, \left(\begin{array}{r} 2\,s-2\\s-1\end{array}
\right)\,
 \left(\frac{\alpha}{n^{^{2}}} \right)^{^{2\,\mbox{\tiny $s$}}}\,\,,
\end{equation}
which converges for
\begin{equation}
\alpha \,\,\leq \,\,\alpha\well\,\,=\,\,\frac{\,n^{\2}}{2}\,\,.
\end{equation}
The quaternionioc counterpart of (\ref{wellC}) is given by
\begin{equation}
\label{wellQ}
\Lambda\well(n,\alpha)=\frac{\Delta \qE\well(n,\alpha)}{E\L}=\frac{\qE\well(n+1,\alpha)-\qE\well(n,\alpha)}{E\L}
\end{equation}
and the $\sigma$-ratio by
\begin{eqnarray}
\sigma\well(n,\alpha)&=&
\Lambda\well(n,\alpha)\,/\,\lambda\well(n)\nonumber\\
 & = & 1\,+\, \frac{1}{2\,n+1}\,\sum_{s=1}^{\infty} (-\,1)^{^{\mbox{\tiny $s$}\,+\,1}} \, \dfrac{2}{s}
\, \left(\begin{array}{r} 2\,s-2\\s-1\end{array}
\right)\,\left[\,(n+1)^{^{2\,-\,4\,\mbox{\tiny $s$}}} - n^{^{2\,-\,4\,\mbox{\tiny $s$}}}  \,\right]
 \alpha^{^{2\,\mbox{\tiny $s$}}}\,\,.
\end{eqnarray}
In Fig.\,2, we plot $\sigma\well(n,\alpha)$ as a function of the $s$-th order correction term for different (convergent) values of $\alpha$. For $n=1$, the convergence is guarantee for $\alpha\leq 1$, see Fig.\,2(a). For $n=2$ the upper (convergence) limit is given by $2$, see Fig.\,2(b).

\subsection*{\normalsize \color{blue!60!black}{IV.C HARMONIC OSCILLATOR}}

The energy eigenvalues for an electron  in the presence of a one-dimensional potential of the form $V(x) = m_e\,\omega^{\2}\, x^2 / 2$, with $\omega$ representing the angular frequency of the oscillator \cite{cohen}, are given by
\begin{equation}
E\osc(n) =\left( n + \frac{1}{2} \right) \, E_{{\omega}}\,\,,
\end{equation}
where $E_{{\omega}} =\hbar \,\omega$. Once known the energy levels, we can calculate the energy gap between sequential energy levels. For the quantum harmonic oscillator, we find
\begin{equation}
\label{oscC}
\lambda\osc(n)=\frac{\Delta E\osc(n)}{E_\omega}=\frac{E\osc(n+1)-E\osc(n)}{E_\omega}=
1\,\,.
\end{equation}
By setting $\abs{W}=E_\omega$ in (\ref{Eseries}), we obtain
\begin{equation}
\frac{\qE\osc(n,\alpha)}{E_\omega} =  n + \frac{1}{2} \, +\,\left( 2\,n + 1 \right)\, \sum_{s=1}^{\infty} (-\,1)^{^{\mbox{\tiny $s$\,+\,1}}} \, \dfrac{1}{s}
\, \left(\begin{array}{r} 2\,s-2\\s-1\end{array}
\right)\,
 \left(\frac{\alpha}{2\,n+1} \right)^{^{2\,\mbox{\tiny $s$}}}\,\,,
\end{equation}
which converges for
\begin{equation}
\alpha \,\,\leq \,\,\alpha\osc\,\,=\,\,n\,+\,\frac{1}{2}\,\,.
\end{equation}
The quaternionioc counterpart of (\ref{oscC}) is given by
\begin{equation}
\label{oscQ}
\Lambda\osc(n,\alpha)=\frac{\Delta \qE\osc(n,\alpha)}{E_\omega}=\frac{\qE\osc(n+1,\alpha)-\qE\osc(n,\alpha)}{E_\omega}\,\,.
\end{equation}
Following the notation used for the hydrogen atom and the potential well, we introduce the $\sigma$-ratio
\begin{eqnarray}
\sigma\osc(n,\alpha)&=&
\Lambda\osc(n,\alpha)\,/\,\lambda\osc(n)\nonumber\\
 & = & 1\,+\, \,\sum_{s=1}^{\infty} (-\,1)^{^{\mbox{\tiny $s$}\,+\,1}} \, \dfrac{1}{s}
\, \left(\begin{array}{r} 2\,s-2\\s-1\end{array}
\right)\,\left[\,(2\,n+3)^{^{1\,-\,2\,\mbox{\tiny $s$}}} - (2\,n+1)^{^{1\,-\,2\,\mbox{\tiny $s$}}}  \,\right]
 \alpha^{^{2\,\mbox{\tiny $s$}}}\,\,.
\end{eqnarray}
In Fig.\,3, we plot $\sigma\osc(n,\alpha)$ as a function of the $s$-th order correction term for different (convergent) values of $\alpha$. For $n=1$, the convergence is guarantee for $\alpha\leq 3/2$, see Fig.\,3(a). For $n=2$ the upper (convergence) limit is given by $5/2$, see Fig.\,3(b).

\subsection*{\normalsize \color{blue!60!black}{IV.D RELATIVISTIC {\small VS} QUTERNIONIC CORRENCTIONS}}
In order to quantify the effect of quaternionic perturbations on bound states, we compare them with relativistic corrections in the hydrogen atom. Relativistic corrections come from the Taylor expansion of the ehergy, 
\[
\sqrt{p^{^{2}}c^{^{2}}\,+\,m_2^{^{2}}c^{^{4}}}\,-\,m_e\,c^{\2}
\,\approx\,\frac{p^{^{2}}}{2\,m_e}\,-\, \frac{p^{^{4}}}{8\,m_e^{^{3}}c^{^{2}}}\,\,.
\]
After some simple algebraic manipulations, we obtain\cite{cohen}
\begin{equation}
\qE\rel(n,\ell) = -\,\frac{R_y}{n^{^{2}}}\, -\,\frac{R_y^{^{2}}}{2\,m_2\,c^{^{2}}n^{^{4}}}\,\left(\,\frac{8\,n}{2\,\ell+1}\,-\,3\,\right)\,\,,
\end{equation}
where $R_y=13.6\,\,\mathrm{eV}$. The energy expansion for a constant quaternionic perturbation $j\,\alpha\,W$ is given by
\begin{equation}
\qE\hyd(n,\alpha\,\abs{W}) = -\,\frac{R_y}{n^{^{2}}}\, -\,\frac{R_y}{n^{^{2}}}\, \sum_{s=1}^{\infty} (-\,1)^{^{\mbox{\tiny $s$\,+\,1}}} \, \dfrac{2}{s}
\, \left(\begin{array}{r} 2\,s-2\\s-1\end{array}
\right)\,
 \left(\frac{n^{\2}\,\alpha\,\abs{W}}{2\,E_y} \right)^{^{2\,\mbox{\tiny $s$}}}\,\,.
\end{equation}
We immediately see that a quaternionic perturbation of $0.15$\,eV is comparable with relativistic correction of the ground state. In the following table, we give the quaternionic corrections for the higher energy levels
\[
\begin{array}{rcrcrcrcrcrc}
 \{\,E\hyd(1) & , & \qE\rel(1,0) & , &  \qE\hyd(1,0.15\,\mathrm{eV}) \,  \} & = &
 \{\,-\,13.60000 & , & -\,13.60090 & , &  -\,13.60080 \,  \}\,\,,\\
 \{\,E\hyd(2) & , & \qE\rel(2,0) & , &  \qE\hyd(2,0.15\,\mathrm{eV}) \,  \} & = &
 \{\,-\,3.40000 & , & -\,3.40015 & , &  -\,3.40331 \,  \}\,\,,\\
 \{\,E\hyd(3) & , & \qE\rel(3,0) & , &  \qE\hyd(3,0.15\,\mathrm{eV}) \,  \} & = &
 \{\,-\,1.51111 & , & -\,1.51116 & , &  -\,1.51854 \,  \}\,\,,\\
 \{\,E\hyd(4) & , & \qE\rel(4,0) & , &  \qE\hyd(4,0.15\,\mathrm{eV}) \,  \} & = &
 \{\,-\,0.85000 & , & -\,0.85002 & , &  -\,0.86313 \,  \}\,\,,\\
\{\,E\hyd(5) & , & \qE\rel(5,0) & , &  \qE\hyd(5,0.15\,\mathrm{eV}) \,  \} & = &
 \{\,-\,0.54400 & , & -\,0.54401 & , &  -\,0.56430 \,  \}\,\,.\\
\end{array}
\]
In Fig.\,4, we show the quaternionic perturbations on the hydrogen atom bound states for different quaternionic potentials. Observe that the upper limit of the quaternionic potential, $\abs{E\hyd(n)}$, is imposed by the convergence condition.

\section*{\large \color{blue!60!black}{V. CONCLUSIONS}}

In this paper, by using the well-established ideas behind time-independent perturbation theory of complex quantum mechanics \cite{cohen}, we have presented a different approach to study the effect of the quaternionic potentials in 
quantum mechanics, i.e. the general solution of the energy eigenvalues in the case of a constant quaternionic perturbations.

Interesting details arise in the calculation of the corrections for the energy eigenvalues of perturbed systems. Such terms show an interesting pattern: those corresponding to odd orders of approximation are null. Hence, the results do not depend on the sign of $\alpha$, since only even powers of the parameter appears. The $j/k$ rotation also gives the same results because in the energy expansion only appears $\abs{W}$. Finally, more important it is the fact that our results differ from what one expects when applying a constant real perturbation in the realm of complex quantum mechanics. For instance, in the complex  scenario, the correction accounts for an equal shift in every energy level of the system.

To provide practical predictions, the general results were applied to some particular examples and the gaps between consecutive energy levels were calculated; the provided graphs show how measurements may detect these quaternionic effects. Of the systems considered, two of them are associated to more relevant topics: the hydrogen atom and the quantum harmonic oscillator.

 The quaternionic perturbation on the quantum harmonic oscillator, it represents a preliminary effort towards a more elaborate model of the system. Some quaternionic extensions of the quantum harmonic oscillator have been proposed, but there does not exist a consensus about which one is more plausible. In \cite{adler}, Adler proposes a fairly natural version, but we lack of mathematical tools to analytically solve the resulting quaternionic differential equation. At fundamental principles, a quaternionic formulation of the quantum harmonic oscillator is essentially important to provide us a better understanding about the construction of quaternionic (quantum) field theories.

We hope that the present work may give new insights about the discussed problems in quaternionic quantum mechanics and contribute to the overall debate.

\section*{\normalsize \color{blue!60!black} \textbf{Acknowledgments}}
The authors thank CNPq (S.D.L./C.A.A.S.) and Fapesp (S.D.L.) for the financial support. One of them (S.D.L.) also thanks the Salento University (Lecce, Italy) for the hospitality during the finalization of the paper.

\newpage

\begin{figure}
\centering
\includegraphics[width=0.8\textwidth]{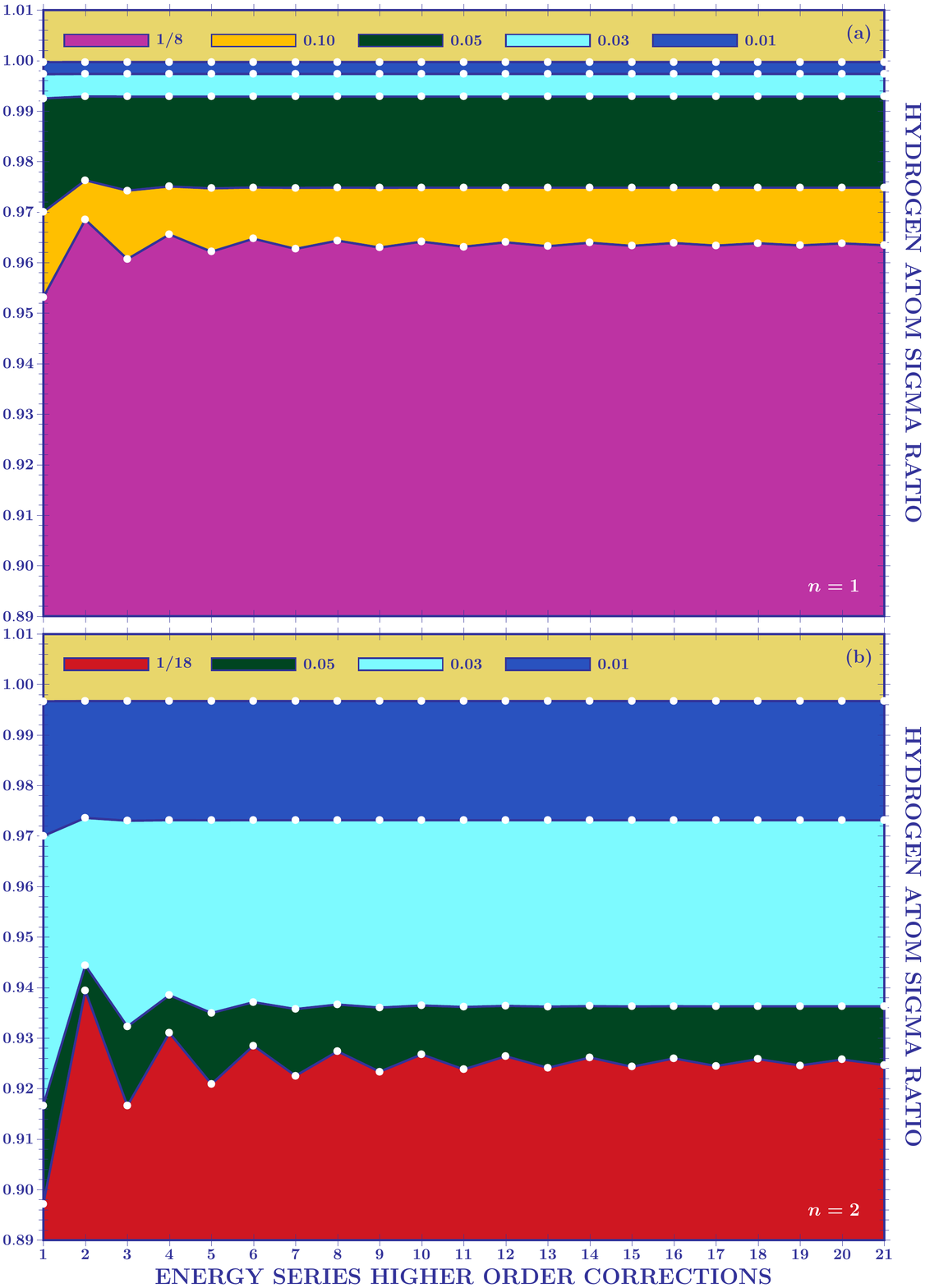}
\caption{The sigma ratio of the hydrogen atom is plotted by using the energy eigenvalue series for different order corrections. The upper convergence limit for $\alpha$ is given by $1/2\,(n+1)^{^{2}}$.}
\end{figure}

\newpage

\begin{figure}
\centering
\includegraphics[width=0.8\textwidth]{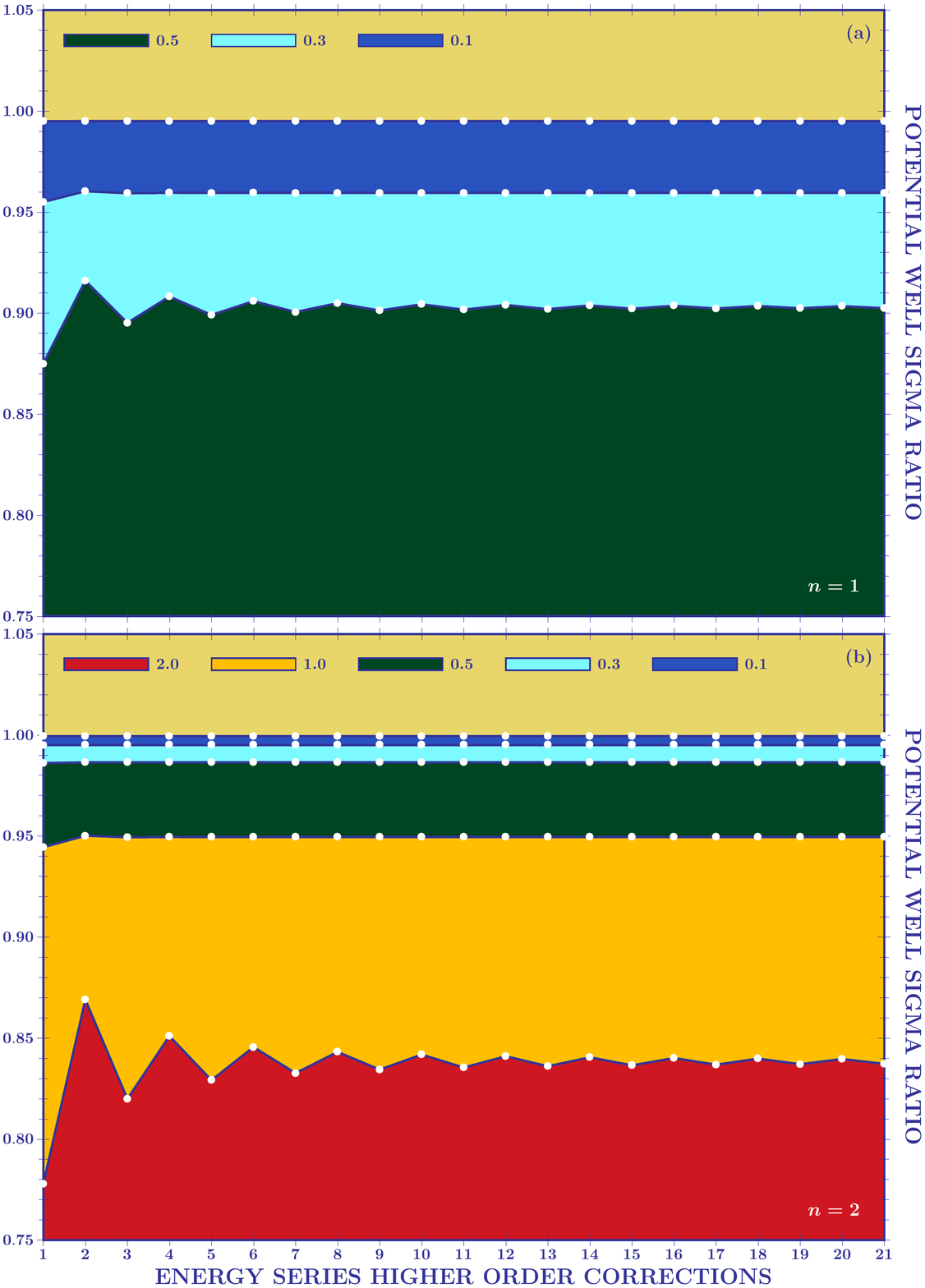}
\caption{The sigma ratio of infinite potential well is plotted by using the energy eigenvalue series for different order corrections. The upper convergence limit for $\alpha$ is given by $n^{^{2}}/2$.}
\end{figure}

\newpage

\begin{figure}
\centering
\includegraphics[width=0.8\textwidth]{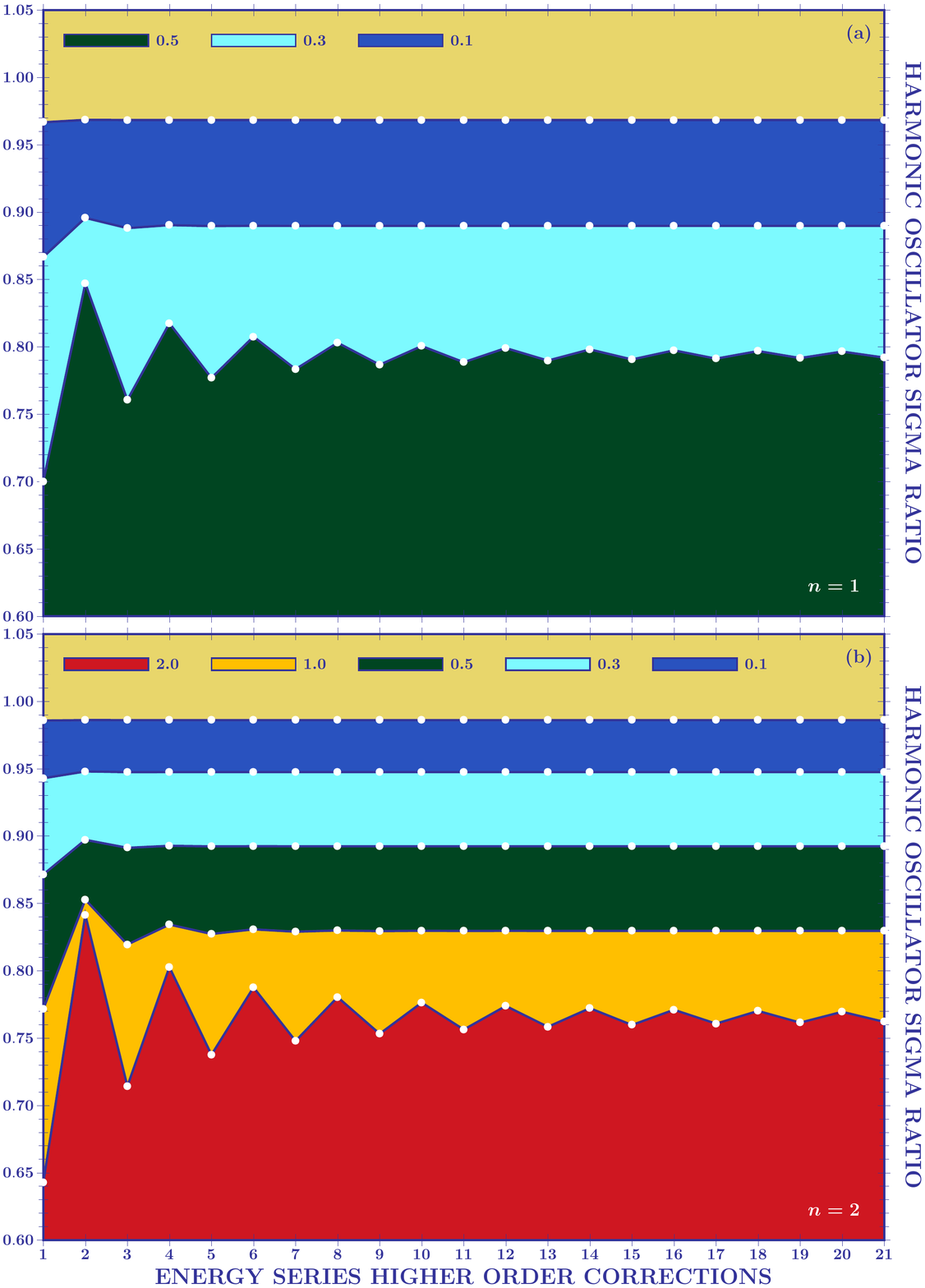}
\caption{The sigma ratio of harmonic oscillator is plotted by using the energy eigenvalue series for different order corrections. The upper convergence limit for $\alpha$ is given by $n+1/2$.}
\end{figure}

\newpage

\begin{figure}
\centering
\includegraphics[width=0.8\textwidth]{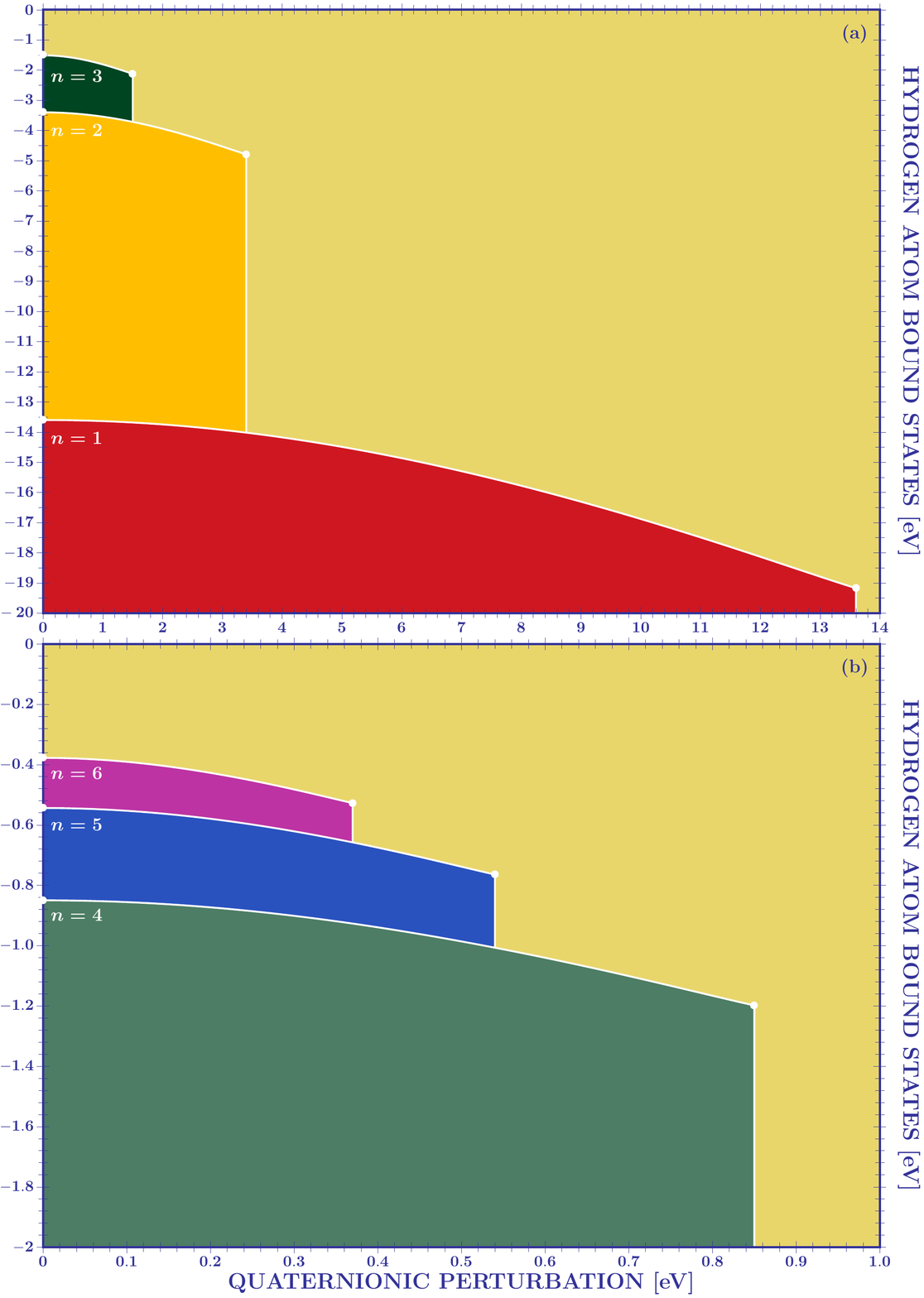}
\caption{Quaternionic perturbations for the first energy levels of the hydrogen atom. The upper limit of the quaternionic potential, $E_y/n^{^{2}}$, guarantees the convergence of the energy eigenvalues series.}
\end{figure}

\end{document}